\input epsf.tex

\documentstyle[aps,twocolumn]{revtex}
\def\An{{\, \rm \AA}}

\def\momt{Q}
\def\meas{{\left[{d \Sigma}/{d 
\Omega}\right]}}
\def\molratio{{x_k}}
\def\pedexp{{m}}

\def\Curtop2001{{Halgren and Damm, 2001}}
\def\Curt2000{{Sheinerman et al., 2000}}
\def\Itri91{{Itri and Amaral, 1991}}
\def\Chen87{{Chen and Lin, 1987}}
\def\parsegian96{{Parsegian and Evans, 
1996}}
\def\Kozin1997{{Kozin et al., 1997}}
\def\Chacon1998{{Chac\'on et al., 1998}}
\def\Ab90{{Abis et al., 1990}}
\def\Hansen86{{Hansen and Mc Donald, 1986}}
\def\ve98{{Velev et al., 1997}}
\def\Baldini99{{Baldini et al., 1999}}
\def\Jac76{{Jacrot, 1976}}
\def\Ash67{{Ashcroft and Langreth, 1967}}
\def\SC98{{Spinozzi et al., 1998}}
\def\Mariani{{Mariani et al., 2000}}
\def\ku97{{Kuehner et al., 1997}}
\def\Vervey48{{Vervey and Overbeek, 1948}}
\def\Israel92{{Israelachvili, 1992}}
\def\Blum78{{Blum and Hoye, 1978}}
\def\Ginoza90{{Ginoza, 1990}}
\def\Hayt81{{Hayter and Penfold, 1981}}
\def\Nagele96{{N\"{a}gele, 1996}}
\def\Hayter82{{Hansen and Hayter, 1982}}
\def\Rog84{{Rogers and Young, 1984}}
\def\GF55{{Guinier and Fournet, 1955}}
\def\Zer86{{Zerah and Hansen, 1986}}
\def\Wagner91{{Wagner et al., 1991}}
\def\Krause91{{Krause et al., 1991}}
\def\DAguann92a{{D'Aguanno and Klein, 1992}}
\def\DAguanno92b{{D'Aguanno et al., 1992}}
\def\Nagel93{{N\"{a}gele et al., 1993}}
\def\Meeron58{{Meeron, 1958}}
\def\JZ81{{Jacrot and Zaccai, 1981}}
\def\gast_mars{{Gasteiger and Marsili, 
1980}}

\def\1qg5internalreference{{Oliveira et al., 
2001}}
\def\Asakura54{{Asakura and Oosawa, 1954}}
\def\Wander94{{Wanderlingh et al., 1994}}
\def\Svergun98{{Svergun et al., 1998}}
\def\Ruiz90{{Ruiz-Estrada et al., 1990}}

\begin{document}

\title{Interaction of proteins in solution\\
  from small angle scattering: a perturbative approach}
\author{Francesco Spinozzi $^1$, Domenico Gazzillo$^2$, Achille
  Giacometti$^2$, Paolo Mariani$^1$ and Flavio Carsughi$^3$}

\address{$^1$ Istituto di Scienze Fisiche, 
Universit\`{a} di Ancona, and \\
INFM Unit\`{a} di Ancona, Via Brecce 
Bianche, I-60131 Ancona, Italy\\
$^2$Dipartimento di Chimica Fisica, 
Universit\`{a} di Venezia, and \\
INFM Unit\`{a} di Venezia, S.Marta 2137, I-
30123, Venezia, Italy\\
$^3$Facolt\`{a} di Agraria, Universit\`{a} 
di Ancona, and \\
INFM Unit\`{a} di Ancona, Via Brecce 
Bianche, I-60131 Ancona, Italy}

\date{\today}
\maketitle

\begin{abstract}
  In this work, an improved methodology for studying interactions of
  proteins in solution by small-angle scattering, is presented. Unlike
  the most common approach, where the protein-protein correlation
  functions $g_{ij}(r)$ are approximated by their zero-density limit
  (i.e. the Boltzmann factor), we propose a more accurate
  representation of $g_{ij}(r)$ which takes into account terms up to
  the first order in the density expansion of the mean-force
  potential. This improvement is expected to be particulary effective
  in the case of strong protein-protein interactions at intermediate
  concentrations.  The method is applied to analyse small angle X-ray
  scattering data obtained as a function of the ionic strength (from
  $7$ to $507$ mM) from acidic solutions of $\beta $-Lactoglobuline at
  the fixed concentration of 10~$\rm g \, L^{-1}$. The results are
  compared with those obtained using the zero-density approximation
  and show a significant improvement particularly in the more
  demanding case of low ionic strength.

\noindent {\sl Running Title}: Interaction of proteins by SAS

\noindent {\sl Keywords}: long-range interactions, mean-force potential,
density expansion, pair correlation functions, 
structure factor,  $\beta$-Lactoglobuline

\end{abstract}
\vskip 0.5cm

\section{Introduction}

\label{introduction}

The study of protein-protein interactions in solution and the
determination of both the physical origin of long range interactions
and the geometry and energetics of molecular recognition can provide
the most effective way of correlating structure and biological
functions of proteins.  In recent years, a large effort has been
devoted to improve the understanding of interactions between
macromolecules in solution. In particular, it has been widely
recognized that the evaluation of electrostatic potentials can produce
quantitative predictions and that factors such as self-energy,
polarizability and local polarity can be biologically crucial
%\cite{Curtop2001,Curt2000}
(\Curtop2001; \Curt2000).  
Nevertheless, major conceptual and
practical problems still exist, and concern, for instance, the
experimental techniques required to measure interaction potentials
under physiologically relevant conditions, as well as the a clarification of
the role of the solvent and of the protein shape and charge
anisotropy.

Several biophysical methods can be used for extracting quantitative
data on protein-protein interactions, even if a detailed analysis of
the long-range interactions has been so far limited to few associating
colloids (\Chen87; \Itri91) and has usually been based on light
scattering or osmotic stress methods
%\cite{parsegian96}
(\parsegian96). However, small angle scattering (SAS) is certainly the
most appropriate tool for studying the whole structure of protein
solutions, because of the small perturbing effects on the system and
the possibility of deriving information on the structural properties and
interactions under very different experimental conditions (pH, ionic
strength, temperature, presence of cosolvents, ligands, denaturing
agents and so on).

In most analyses of SAS data, particle interactions are however
disregarded, 
assuming either large separation or weak interaction forces. 
The interactions among macromolecules determine their spatial
arrangement, which can be described by correlation functions.  These
functions
may be related, for instance via integral equations, to the
 direct pair
potentials, describing the interaction between two particles. 
When the average distance among particles 
is large or the interaction potentials are weak, 
the influence of the average structure factor of the system
(i.e. the Fourier transform of the average correlation function) may
be negligible inside the considered experimental angular window, and 
the particles can be reckoned as completely uncorrelated.  Under these
conditions, the SAS intensity appears to depend only upon the average form
factor. Note that this approximation of neglecting all intermolecular 
forces is used in most applications of X-ray or neutron SAS
(\Kozin1997; \Chacon1998).

When the above conditions are not verified, then particles 
cannot be considered uncorrelated, and the average
structure factor cannot be neglected in the expression of the SAS
intensity. In this case data analysis is far more complicate.
In principle, asymptotic behaviors could be used to separate the SAS 
intensity into (average) form and structure factors
 (\Ab90). 
If the particle form factors are known, an experimental average structure
factor can be extracted, by dividing the intensity
by the average form factor. Then, some insight into the intermolecular
forces may be obtained by comparison with the theoretical
structure factor calculated from some interaction
model, by using analytical or numerical methods from the statistical
mechanical theory of liquids (\Hansen86).

Unfortunately, the most powerful and accurate techniques provided by
this theory - such as Monte Carlo and molecular dynamics computer
simulations as well as integral equations - can hardly be included
into a typical best-fit procedure for analysing experimental data.
Working at very low concentrations, a first possibility of improving over
the crude recipe of neglecting the average structure factor is to
evaluate that quantity  
 by approximating the pair correlation functions 
$g_{ij}(r)$ with their zero-density limit, given by the Boltzmann
 factor (\ve98). 
In the present paper, we shall show that this zero-density
approximation becomes 
quite unusable at the usual protein concentrations when the ionic strength is 
low, i.e., in the presence of strong electrostatic interactions. Clearly, it would be desirable to find an alternative, simple but reasonably accurate, way
for computing the average structure factor of globular proteins at low
or moderate concentrations. This is the major aim of our paper.

Although the new proposal is methodological and thus applicable, in principle,
to a wide class of spherically symmetric interaction models, it will
be illustrated on a concrete case, as a part of a more general study
on structural properties of a particular protein in solution, $\beta$-Lactoglobulin ($\beta$LG).

In a previous paper (\Baldini99), which provides a natural
introduction to the present work, all long-range protein-protein interactions
were neglected and the average structure factor was assumed to be unity.
That investigation reported experimental data concerning
structural properties of $\beta$LG acidic solutions (pH 2.3), at several values of ionic strength in the range 7-507 mM
%\cite{Baldini99}
(\Baldini99).  
Photon correlation spectroscopy and small angle X-ray
scattering (SAXS) experiments gave a clear evidence of a monomer-dimer
equilibrium affected by the ionic strength.  
In the angular region
where SAXS experiments were performed, the contribution of long-range
protein-protein interactions was expected to be rather small.
Accordingly, SAXS data were analysed only in terms of $\beta$LG
monomer and dimer form factors, which were calculated very accurately.
Short-range forces responsible for protein aggregation were taken
 into account only implicitly through a chemical
association equilibrium, employed to evaluate the dimerization
fraction. A global fit procedure allowed the determination of the
monomer effective charge, as well as of the protein dissociation free
energy within a wide range of ionic strength
%\cite{Baldini99}
(\Baldini99).

In the present paper, we shall investigate, within the same physical
system, the long-range protein-protein interactions, which can strongly
influence the small-angle scattering at low ionic strength. 
To this aim, two issues have to be addressed.
First, one needs to extend the experimental SAXS angular region to
lower values of the scattering vector, where long-range forces play an
important role.
Second, one has to select an accurate and tractable theoretical
scheme for calculating the average structure factor to be used in the fit
of experimental data. Both tasks have been accomplished in this work.

We first report a new set of SAXS measurements on $\beta$LG performed
under the same experimental conditions of Baldini and coworkers
%\cite{Baldini99}
(\Baldini99), but for smaller angles.
These data unambiguously display a lowering in the scattering 
intensity at small angles, with a progressive development of
an interference peak, when ionic strength is low.
This occurrence is a clear signal of strong protein-protein interactions,
 and we shall show that it can be simply interpreted in terms of
screened electrostatic repulsions among charge macroions.

Next, we shall propose an improvement for the calculation of the theoretical average structure factor, based upon a new approximation to the protein-protein correlation functions $g_{ij}(r)$. Starting from the density expansion of the corresponding mean-force potentials, we shall show that the simple addition of the $1^{\rm st}$-order perturbative correction to the direct pair potentials leads to a marked progress with respect to the use of the Boltzmann factor, while retaining the same level of simplicity. The new approximation is indeed able to predict, at low ionic strength, the interference peak observed in the experimental scattering intensity, and consequently it leads to a significantly improved fit.

We stress, in advance, that a check of the unavoidable
limits of validity of the proposed approach will not be treated
here. A further study involving a comparison with more accurate
theoretical results (from Monte Carlo or molecular dynamics, as well as from integral equations) is, of course, desirable, but goes beyond the scope of the present paper, and will be left for future work.

\vskip 1 cm

\section{Basic theory}
\label{theory}

Because of the presence of an aggregation equilibrium, a $\beta$LG
solution contains two different forms of macroions (protein monomers
and dimers) embedded in a suspending fluid and in a sea of microions,
which include both counter-ions neutralizing all protein charges and
small ions originated 
from the addition of electrolyte salts. To represent such
a system, we shall employ a simple ``two-component macroion model'',
which effectively takes into account only protein particles. 
Within this scheme, which is usually referred to as the
Derjaguin-Landau-Vervey-Overbeek (DLVO) model
%\cite{Vervey48}
(\Vervey48), the suspending fluid (solvent) is represented as a uniform
dielectric continuum and all microions are treated as point-like
particles. The presence of both solvent and microions appears only
in
the macroion-macroion {\it %
  effective } potentials. A further simplification follows from
the assumption of spherically symmetric interactions. 
We note that in our model,
component 1 and 2 correspond to monomers and dimers, respectively.

Before addressing the specific system under investigation, 
it is convenient to recall some basic points
of the general theory.

\subsection{Scattering functions}

The macroscopical differential coherent scattering cross section
$d\Sigma/d\Omega $, obtained from a SAS experiment, is related to the
presence of scattering centers, i.e. density and/or structural
inhomogeneities, and can yield quantitative information about their
dimensions, concentration as well as shape and interaction potentials.
The cross section is proportional to the ``contrast'', namely the
difference of electron density multiplied by the classical electron
radius (or scattering length density in the neutron case) between the
scattering centers and the surrounding medium; in the case of
biological samples, this quantity can also be tuned in order to obtain
more detailed information about the scattering structures (contrast
variation technique
%\cite{Jac76}
\Jac76). 
Proteins in solution represent an excellent example of
inhomogeneities for SAS measurements, due to their high contrast with
X-rays (as well as with neutrons). The general equation for the SAS
intensity is
\begin{equation}
{\frac{d\Sigma }{d\Omega }}({\bf Q})=\frac 
1V\left\langle \left| \int_Vd{\bf %
r}\delta \rho ({\bf r})e^{{\rm i}{\bf 
Q}\cdot {\bf r}}\right| ^2\right\rangle ,
\label{sas1}
\end{equation}
${\bf Q}$ being the exchanged wave vector, 
with magnitude $%
Q=\left( 4\pi /\lambda \right) \sin {\theta ,}$ where $\lambda $
represents the incident radiation wavelength and $2\theta $ is the
full scattering angle. The integral in Eq.~\ref{sas1} is
extended over the sample volume $V$%
, with ${\bf r}$ being the position vector and $\delta \rho ({\bf r})$
the fluctuation with respect to a uniform value, $\rho_0$, of the
local electron density multiplied by the classical electron radius (or
simply the scattering length density in the case of neutrons). Angular
brackets represent an ensemble average over all possible
configurations of the proteins in the sample.

Eq.~\ref{sas1} can be reduced to a simpler form, when the interactions
are spherically symmetric. Using a ``two-phase'' representation of the
fluid (only one type of homogeneous scattering material with
scattering density $\rho_P$ inside proteins, embedded in a homogeneous
solvent phase with density $\rho_0$) yields
\begin{eqnarray}
{\frac{d\Sigma }{d\Omega }}(Q) &=&(\Delta 
\rho )^2\biggl\{ %
\sum_{i=1}^pn_iV_i^2\left[ <\!F_i^2({\bf 
Q})\!>_{\omega _Q}-<\!F_i({\bf Q}%
)\!>_{\omega _Q}^2\right] \nonumber \\
&&+\sum_{i,j=1}^p
\left( n_i~n_j\right) ^{1/2}
~V_i~V_j~ <\!F_i({\bf Q})\!>_{\omega 
_Q}<\!F_j(%
{\bf Q})\!>_{\omega _Q}S_{ij}(Q)\biggr\} 
\label{iqtot}
\end{eqnarray}
where $\Delta \rho \equiv \rho_P-\rho_0$ represents the contrast, $p$
the number of protein species (2 for our solutions with monomers and
dimers), $n_i$ the number density of species $i$, $V_{i\text{ }}$the
volume, $F_i({\bf Q})$ the form factor, $S_{ij}(Q)$ the Ashcroft-
Langreth partial structure factor and $<\!...\!>_{\omega _Q}$ denotes
an orientational average.

The partial structure factors 
%\cite{Ash67}
(\Ash67) are defined as
\begin{equation}
S_{ij}(Q)=\delta _{ij}+4\pi \left( 
n_i~n_j\right) ^{1/2}\int_0^\infty
dr~r^2~\left[ g_{ij}(r)-1\right] ~\frac{\sin 
(Qr)}{Qr},
\label{sqpart}
\end{equation}
in terms of the three-dimensional Fourier transform of $g_{ij}(r)-1 $,
where $g_{ij}(r)$ is the pair correlation function (or radial
distribution function) between particles of species $i$ and $j.$

Finally, the average form and structure factor, $P(Q)$ and $S_M(Q),$
are
\begin{equation}
P(Q)=(\Delta \rho 
)^2~\sum_{i=1}^p~n_i~V_i^2~<\!F_i^2({\bf 
Q})\!>_{\omega
_Q}, \label{pqeff}
\end{equation}
\begin{equation}
S_M(Q)={\frac{d\Sigma }{d\Omega }}(Q)\ /\ 
P(Q). \label{sqeff}
\end{equation}

\subsection{Protein form factors}

The angular averaged form factor of species $i$ can be written as
\begin{equation}
<\!F_i({\bf Q})\!>_{\omega _Q}=\int_0^\infty 
~dr~p_i^{(1)}(r)~\frac{\sin (Qr)%
}{Qr}, \label{pqeff2}
\end{equation}
where $p_i^{(1)}(r)$ represents the probability for the $i$-th species
that a point at distance $r$ from the protein center of mass lies
inside the macromolecule. Similarly, the angular averaged squared form
factor is given by
(\GF55)

\begin{equation}
<\!F_i^2({\bf Q})\!>_{\omega 
_Q}=\int_0^\infty 
~dr~p_i^{(2)}(r)~\frac{\sin
(Qr)}{Qr} \label{pqeff3}
\end{equation}

\noindent where $p_i^{(2)}(r)$ represents the probability for the
$i$-th species
to find a segment of length $r$ with both ends inside the
macromolecule. Both integrals of $p_i^{(1)}(r)$ and $p_i^{(2)}(r)$ are
normalized to unity. These distribution functions have been calculated
from the crystallographic structures of both the monomer 
and dimer forms of the protein, as described in Refs.
(\Baldini99; \Mariani), briefly recalled in Appendix A, and discussed 
in Subsection III C.

\subsection{Protein-protein interaction potentials}
The choice of the proper potential is a rather delicate matter and depends on
the investigated system. For instance, in a study on lysozyme 
%\cite{ku97}
(\ku97) the
protein-protein interaction was assumed to be the sum of four
contributions, namely a hard-sphere term, an electrostatic repulsion,
an attractive dispersion potential and a short-range attraction. 
In a different study, 
on lysozime and chymotrypsinogen 
%\cite{ve98}
(\ve98) five contributions were, on the other hand, considered: 
charge-charge repulsion,
charge-dipole, dipole-dipole and van der Waals attraction, along 
with further complex short-range interactions. 
In this paper we follow
a different route motivated by the fact that the presence of several
interaction terms may obscure the relative importance of each of them.
Moreover, the choice of a very refined potential would be in striking
contrast with the very crude approximations used in calculating the
RDFs. On this basis we shall search for the simplest possible model
potential which is still capable of capturing the essential features
of the system. It will be the sum of two repulsive contributions:
\begin{equation}
u_{ij}(r)=u_{ij}^{{\rm HS}}(r)+u_{ij}^{{\rm 
C}}(r) \label{potentials}
\end{equation}
where
\begin{equation}
u_{ij}^{{\rm HS}}(r)=\left\{
\begin{array}{ll}
\mbox{$+\infty$} & \quad \mbox{$0\leq 
r<R_i+R_j$} \\
\mbox{0} & \quad \mbox{$r\geq R_i+R_j$} \\
\end{array}
\right.
\end{equation}
is a hard-sphere (HS) term which accounts for the excluded-volume
effects ($R_i$ being the radius of species $i$) and
\begin{equation}
u_{ij}^{{\rm 
C}}(r)=\frac{Z_iZ_je^2}{\varepsilon 
(1+\kappa _DR_i)(1+\kappa
_DR_j)}\ \frac{\exp [-\kappa _D(r-R_i-
R_j)]}r
\end{equation}
represents a screened Coulomb repulsion between the macroion charges,
which are of the same sign. This term has the same Yukawa form as in
the Debye-H\"{u}ckel theory of electrolytes, but the coupling
coefficients are of DLVO type
%\cite{Vervey48}
(\Vervey48).  Here, $e$ is the elementary charge, $\varepsilon $ the
dielectric constant of the solvent and the effective valency of
species $i$, $Z_i,$ may depend on the pH. The inverse Debye screening
length $\kappa _D$, defined as
\begin{equation}
\kappa _D=\left[ \frac{8\pi \beta 
e^2N_A}\varepsilon \left( I_S+I_{{\rm c}%
}\right) \right] ^{1/2}, \label{kdebye}
\end{equation}
depends on temperature ( $\beta =(k_BT)^{- 1}$ ) and on the ionic
strength of all microions. $I_S$ and $I_{{\rm c}}$ represent the ionic
strength of all added salts ($S$) and of the counterions ($c$),
respectively. Both these terms are of the
form $(1/2)\sum_ic_i^{{\rm micro}%
  }(Z_i^{{\rm micro}})^2$, with $c_i^{{\rm micro}}=n_i^{{\rm
    micro}}/N_A$ being the molar concentration of micro- species $i$
($N_A$ is Avogadro's number). $I_{{\rm c}}$ is related to the macroion
number densities $n_1$ and $n_2$ (1 = monomer, 2 = dimer) through the
electroneutrality condition, according to which the counterions must
neutralize all macroion charges, i,.e. $n_{{\rm c}}\left| Z_{{\rm
      c}}\right| =n_1\left| Z_1\right| +n_2\left| Z_2\right| $. Notice
that the dependence of $\kappa _D$ on $I_S$ implies that the strength
of the effective potential $u_{ij}^{{\rm C}}(r)$ can largely be varied
by adding an electrolyte to the solution.

We have explicitly checked that the addition of an attractive term
with the form of a Hamaker potential $u_{ij}^{{\rm H}}(r)$
%\cite{Israel92}
(\Israel92) does not alter our final conclusions. The basic reason for
this can be traced back to the fact that van der Waals attractions
may be
completely masked by $u_{ij}^{{\rm C%
    }}(r),$ when the electrostatic repulsion is strong, and are also
negligible for moderately charged particles with
diameter smaller than 50 nm %\cite{Nagele96}
(\Nagele96).  Moreover, $u_{ij}^{{\rm H}}(r)$ diverges at $r=R_i+R_j$,
so that its applicability could be preserved only by the addition of a
non-interpenetrating hydration/Stern layer
%\cite{Baldini99,ku97}
(\Baldini99; \ku97).

We stress the fact that some attractive interactions must, however, be
present in the system, since they are responsible for the aggregation
of monomers into dimers, and determine the value of the monomer molar
fraction  $x_1$,  which is required to complete the definition of our
model. However, due to the complexity of these interactions (including
hydrogen bonding), a clear understanding of their explicit functional forms
is still lacking. Therefore, following Baldini et al. (1999), we will
account for them indirectly, by using a chemical association
equilibrium to fix  $x_1$. 
The dissociation free energy, which determines the equilibrium
constant, is written as a sum of two contributions, i.e.
\begin{eqnarray}
\Delta G_{\mathrm{dis}} &=& \Delta 
G_{\mathrm{el}}+\Delta G_{\mathrm{nel}},
\end{eqnarray}
where $\Delta G_{\mathrm{el}}$ is an electrostatic term calculated
within a Debye-H\"uckel theory, and $\Delta G_{\mathrm{nel}}$ is an
unknown non-electrostatic contribution, which will be left as a free
parameter in the best-fit analysis.

\subsection{Radial distribution functions}

Given a model potential, one has to calculate the corresponding radial
distribution functions (RDF) $g_{ij}(r),$
which can be expressed by the {\it %
exact} relation
\begin{equation}
g_{ij}\left( r\right) =\exp \left[ -\beta 
W_{ij}\left( r\right) \right] ,
\label{rad1}
\end{equation}

\begin{equation}
-\beta W_{ij}\left( r\right) = -\beta u_{ij}\left( r\right) 
+\omega _{ij}(r)
\end{equation}
where $W_{ij}\left( r\right)$  is the potential of mean
 force, which includes the direct pair
potential $u_{ij}\left( r\right) $ as well as $-\beta^{-1}
 \omega_{ij}(r)$, i.e.
the indirect interaction between $i$ and $j$ due to their interaction
with all remaining macroparticles of the fluid. In the zero-density
limit, $\omega _{ij}(r)$ vanishes and $g_{ij}\left( r\right) $ reduces
to the Boltzmann factor, i.e.
\begin{equation}
g_{ij}\left( r\right) =\exp \left[ -\beta 
u_{ij}\left( r\right) \right]
\qquad \text{as~} n \rightarrow 0, 
\label{boltz}
\end{equation}
which represents a $0^{\rm th}$-order approximation, frequently used
in the analysis of experimental scattering data ($n \equiv \sum_m n_m$
is the total number density).

The most common procedure for determining an accurate $g_{ij}(r)$ or,
equivalently, the correction term $\omega _{ij}(r)$, would be to solve
the Ornstein-Zernike (OZ) integral equations of the liquid state
theory, within some approximate closure relation
%\cite{Hansen86}
(\Hansen86). This can typically be done numerically, with the
exception of few simple cases (for some potentials and peculiar
closures) where the solution can be worked out analytically.

For our hard-sphere-Yukawa potential (neglecting the Hamaker term),
the OZ equations do admit analytical solution, when coupled with the
``mean spherical approximation'' (MSA)
%\cite{Blum78,Ginoza90,Hayt81}
(\Blum78; \Ginoza90; \Hayt81).  Nevertheless, at low density and for
strong repulsion the MSA RDFs may assume unphysical negative values
close to interparticle contact
%\cite{Nagele96}
(\Nagele96). To overcome this difficulty, it would be possible to
utilize an analytical ``rescaled MSA''
%\cite{Nagele96,Hayter82}
(\Nagele96; \Hayter82;  \Ruiz90 ), 
or to resort to different closures
(Rogers-Young approximation or ``hypernetted chain'' closure), which
compel numerical solution
%\cite{Rog84,Zer86,Wagner91,Krause91,DAguann92a,DAguanno92b,Nagel93}
(\Rog84; \Zer86; \Wagner91; \Krause91; \DAguann92a; \DAguanno92b;
\Nagel93).

In more general, when only numerical solutions are feasible, integral
equation algorithms can hardly be included in a best-fit program for
the analysis of SAS results.  The use of
analytical solutions, or simple approximations requiring only a minor
computational effort, is clearly much more advantageous when fitting
experimental data.  The $0^{\rm th}$-order approximation
given in Eq.~\ref{boltz} avoids the problem of solving the OZ
equations, but is largely inaccurate except, perhaps, at
very low densities.

In order to improve over this $0^{\rm th}$- order approximation to the
RDFs, the basic idea put forward in the present work hinges upon the
expansion of the potential of mean force into a power series of the
total number density $n$
%\cite{Meeron58}
(\Meeron58). Neglecting all terms beyond the first order,
Eq.~\ref{rad1} then becomes
\begin{equation}
g_{ij}\left( r \right) =\exp 
\left[ -\beta u_{ij}\left( r\right) +\omega
_{ij}^{(1)}(r)n
\right].
\label{1st}
\end{equation}

By construction, this expression 
is never negative, thus avoiding the major drawback of MSA. 
The explicit expression for the
perturbative correction $\omega _{ij}^{(1)}(r)$ is given in Appendix
B. The considered $1^{\rm st}$-order approximation substantially improves the
accuracy of the RDFs with respect to Eq.  \ref{boltz}, while remaining
at nearly the same level of simplicity (see Appendix B). Moreover, it
is to be stressed that the usage of the new approximation is not
restricted to the model of this paper, but the proposed calculation
scheme can be equally well applied to different spherically symmetric
potentials.

\section{Materials and methods}

\subsection{Samples}

A bovin milk $\beta$LG B stock solution (concentration 40 ${\rm g \,
  L^{-1}}$) was obtained by ionic exchange of protein samples against
a 12 mM phosphate buffer (ionic strength $I_{S}= 7\ {\rm mM}$ and
${\rm pH=2.3}$)
%\cite{Baldini99}
(\Baldini99).  Nine samples at ionic strength 7, 17, 27, 47, 67, 87,
107, 207, 507 mM were then prepared by adding appropriate amounts of
NaCl.  The final protein concentrations were about 10 ${\rm g \,
  L^{-1}}$.

The monomeric $\beta$LG unit is composed by 162 amminoacid residues
and has a molecular weight of $18400\ {\rm Da}$.  The excluded protein
volume has been calculated from the amino acid volumes, as reported by
Jacrot and Zaccai
%\cite {Jac76,JZ81}
(\Jac76; \JZ81). The monomer volume results to be $V_{1}=23400{\ {\rm
    \AA }}^{3}$; hence, the $\beta$LG electron density is
$\rho_P=0.418 \, \rm e \AA^{-3}$.
By considering the basicity of the amino acids, at ${\rm pH=2.3}$ the
monomer charge would be near $20e$.  This result is confirmed by the
Gasteiger- Marsili method
%\cite{gast_mars}
(\gast_mars), assuming that all amino groups 
${\rm %
  NH_2}$ are protoned at ${\rm pH=2.3}$.  The crystallographic
structure of $\beta$LG both in monomer and in dimer form can be found
in the Protein Data Bank,
%\cite{PDB}
%(\PDB), 
entry 1QG5
%\cite{1qg5internalreference}
(\1qg5internalreference). A sketch of $\beta$LG dimer structure can be
found in Fig.~1 of Ref.
%\cite{Baldini99}
(\Baldini99).  It can be observed that all 20 basic amino acids are on
the protein surface, but two of them are at the monomer-monomer
interface;
therefore at ${\rm pH=2.3}$ the ratio $%
Z_2/Z_1$ between dimer and monomer charges could be about $1.8$.

\subsection{SAXS experiments}

SAXS measurements were collected at the Physik Department of the
Technische Universit\"at M\"unchen (Germany) using a rotating-anode
generator. The radiation wavelength was $\lambda =0.71{\ {\rm \AA }}$
and the temperature $20 ^{\circ} \mbox{C} $.  The $Q$ range was
$0.035-0.1 {\ {\rm \AA}}^{-1}$.  $\beta$LG samples were measured in
quartz capillaries with a diameter of $2\ {\rm mm}$ and a thickness of
$10\ \mu {\rm m}$ (Hilgenberg, Malsfeld, D). X-ray patterns were
collected by a two-dimensional detector and radially averaged.  The
scattering from a solvent capillary was subtracted from the data after
correction for transmission, capillary thickness and detector
efficiency.

\subsection{Best-Fit analysis}

A previous analysis of SAXS data for similar samples in the range
$Q=0.07 \div 0.3 {\, {\rm \AA}}^{-1}$ has been recently reported by
some of us
%\cite{Baldini99}
(\Baldini99). In the present work we have 
extended these experiments
to the range $%
Q=0.035 \div 0.1 {\, {\rm \AA}}^{-1}$, 
where protein-protein interactions are expected to play a major role. 
The two sets have then been combined into a single set of measurements
 with $Q$ ranging from $0.035$ to $0.3 {\, {\rm \AA}}^{-1}$.

As regards the calculation of the monomer and dimer form factors, it is well
 known that the scattering form factor of a biomolecule in solution 
depends on the crystallographic coordinates and the form factors of
all constituent atoms, as well as on the hydration shell of the
 resulting macroparticle.  Computer programs such as CRYSOL (Svergun
 et al., 1995) are able to calculate such a form factor, taking all
 the above-mentioned variables into account.
 It is also widely accepted that the SAS technique is a low-resolution
 one, and approximating the $\beta$LG protein by a homogeneous
 scattering particle yields comparable results up to $Q= 0.4
 {\, {\rm \AA}}^{-1}$,  as we have tested by checking our method
 against the results of the CRYSOL software.
  The equivalent homogeneous scattering particle has a shape defined
 by the envelope of the van der Waals spheres centered on each atom.
 The SAS community often exploits the Monte Carlo method to calculate
 the form factor of a given shape (Henderson, 1996). We have modeled
 the hydration shell with a semigaussian function, instead of a linear
 one proposed by Svergun (Svergun et al., 1997). Our simple and efficient
 method has already been applied with success in previous works
 (Baldini et al., 1999; Mariani et al., 2000).

The Monte Carlo method used to calculate the distribution functions
 $p^{(1)}_i(r)$ and $p^{(2)}_i(r)$ of both
monomers ($i=1$) and dimers ($i=2$) from their
crystallographic structures is outlined in
Appendix A.  Then the form factors $<\! F_{i}({\bf Q})
\!>_{\omega_{Q}}$ and $<\! F_{i}^2({\bf Q}) \!>_{\omega_{Q}}$ have
been obtained through Eqs.~\ref{pqeff2} and \ref{pqeff3}, by
calculating the radial integrals with a grid size of $1 \An$ up to a
maximum $r$ corresponding to $p^{(i)}(r)=0$, $(i=1,2)$.

According to the dissociation free energy model described in Ref.
%\cite{Baldini99}
(\Baldini99), the monomer molar fraction $x_1$ is a function of the
ionic strength $I_S$. This suggests the possibility of a simultaneous
fit for all SAXS intensities curves, using just few parameters, all
independent of $I_S$.  In particular, as in Baldini {\it et al.}
%\cite{Baldini99}
(\Baldini99), the following parameters have been fixed:  
 the dielectric constant of the solvent,
$\varepsilon=78.5$; the experimental temperature, $T=293~ {\rm K}$;
the ratio between the effective charges of dimer and monomer,
$Z_2/Z_1=1.8$; the monomer
and dimer ``bare'' radii, $R_{1} =19.15 {\ {\rm \AA}}$ and $R_{2}
=2^{1/3} R_{1}$.
The choice for $R_{2}$ is easily understood if we recall that 
 our model of long-range
 interactions involves the approximation of considering a dimer as a
sphere with volume twice as large as the monomer one. This
introduction of an equivalent sphere is a simplifying
 approximation often used by the SAS community. On the other hand, 
we have calculated the form factor of the dimer from its exact,
 rather elongated form. 

In the global fit the only free parameters are
therefore $Z_1$ and $\Delta G_{\mathrm{nel}}$, the non-electrostatic
free energy. The merit functional to be minimized was defined as
\begin{eqnarray}
\chi^2&=& \frac{1}{N_S} \sum_{\pedexp=1}^{N_S} 
{\bar \chi}^2_\pedexp \nonumber \\
{\bar \chi}^2_\pedexp &=& \frac{1}{N_{\momt,\pedexp}} 
\sum_{i=1}^{N_{\momt,\pedexp}} \left\{
\frac{{\
\meas_\pedexp^{exp} (\momt_i)- \kappa_\pedexp 
\meas_\pedexp^{fit} (\momt_i)-B_\pedexp} }
{\sigma_{\pedexp}(\momt_i) }\right\}^2
\label{partial}
\end{eqnarray}

\noindent where $N_S$ is the number of scattering curves under analysis,
$N_{Q,\pedexp}$ is the number of experimental points in the $\pedexp-
$th curve, and $%
\sigma_\pedexp(Q_i)$ is the experimental uncertainty on the intensity
value at $Q_i$.  $\meas_\pedexp^{fit}(\momt_i)$ is the corresponding
cross section predicted by the model by using Eq.~\ref{iqtot}; for
each experiment, the calibration factor $\kappa_\pedexp$ and the flat
background $B_\pedexp$ have been adjusted from a linear least-squares
fit of $\meas_\pedexp^{exp}(Q)$.  The partial structure factors,
Eq.~\ref{sqpart}, have been calculated with an integration upper limit
of $r=500 \An$ and a grid size of $1 \An$.

The physical meaning of the ``flat background'' requires a comment, 
since constant subtraction is usually accepted for neutron scattering,
 but not for X-ray scattering.
Introducing these backgrounds is suggested by observing that 
 one of major experimental problems with X-rays is the exact
 determination of the transmission factor. A non-exact value
 would result into a non-perfect subtraction of the background due
 to the electronic noise.
However, as shown later in Table II, the low values obtained
 for $B_\pedexp$, as compared to the values of the scaling
factors, indicate that these parameters play a minor role in the
 data analysis.

  Typical calculation times for the best-fit 
on a Digital Alpha 433 are a few minutes for the $0^{\rm th}$-order
approximation and $\simeq 20$ hours for the $1^{\rm st}$-order one.
The effect of experimental errors on the fitting parameters has been
determined using a sampling method.  For each scattering curve, we
start from $N_{Q,\pedexp}$ intensities $\meas_\pedexp^{exp} (\momt_i)$
with their experimental standard deviation and we generate $N_I$ new
data sets (for $\beta$LG we used $N_I =15$) by sampling from
$N_{Q,\pedexp}$ gaussians of width $\sigma_\pedexp(Q_i)$ centred at
the observed values.  Each data set generated for all curves is
then analyzed with the global fit algorithm described earlier.  The
errors on the fitting parameters, $Z_1$ and $\Delta G_{\mathrm{nel}}$,
and on the scaling parameters, $\kappa_\pedexp$ and $B_\pedexp$, are
obtained by calculating their values from each data set and, finally,
their standard deviation from the first value.

\section{Results and Discussion}

\begin{figure}[tbp]
\centerline{\epsfxsize=3.5truein \epsfysize=3.5truein
\epsffile{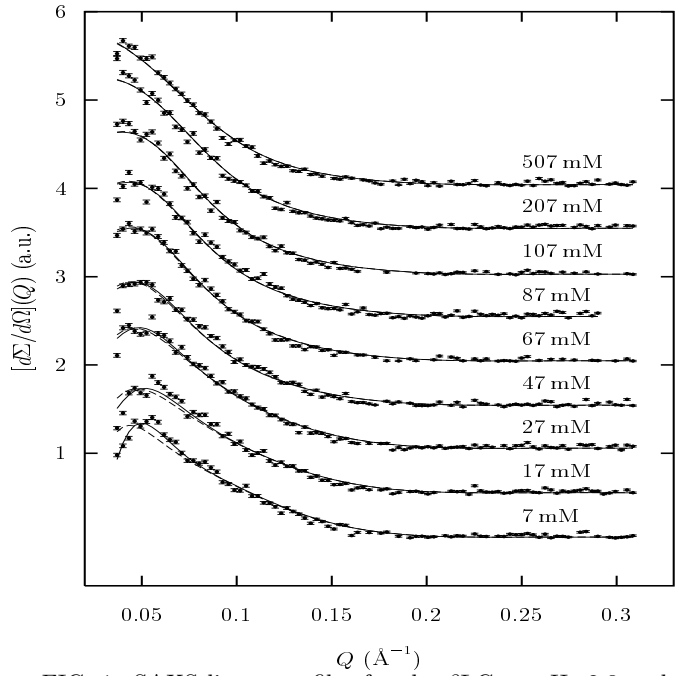} }
\caption{SAXS linear profiles for the $\beta$LG at 
  pH=2.3 and concentration 10~$\rm g \, L^{- 1}$ in different ionic
  strength conditions (as indicated above each curve).  Points are
  experimental results, whereas the dashed and the solid lines
  represent the best fits obtained by applying the $0^{\rm th}$-order
  and $1^{\rm st}$- order approximations of the pair correlation
  functions, respectivley.  The curves are scaled for clarity by a
  factor $0.5$.}
\label{figure1}
\end{figure}

Fig.~\ref{figure1} depicts the experimental results for the X-ray
intensity $\meas(Q)$ as a function of the transferred momentum $Q$ at
several values of ionic strength.
Here, instead of the usual logarithmic scale, we have 
preferred the use of a linear scale, in order to let the reader
appreciate more easily the small differences between experimental data
and theoretical curves. On a log scale these differences would be
hardly visible.

Our measurements 
clearly show the formation and evolution of an interference peak at
 small angles, as the ionic strength decreases.  The appearance of 
such a peak is evidently due to increasing protein-protein interactions.
In the same figure, the performance of
our $1^{\rm st}$-order approximation is compared with that of the
commonly used $0^{\rm th}$-order one. The $1^{\rm st}$- order
approximation yields a fit of rather good quality through the whole
measured range $Q$.  The development of the interference peak,
underestimated by the $0^{\rm th}$-order approximation, is now well
reproduced, indicating that the main physical features of the
$\beta$LG solution are indeed taken into account by our simple
interaction model.

\begin{figure}[tbp]
\centerline{\epsfxsize=3.5truein \epsfysize=3.5truein
\epsffile{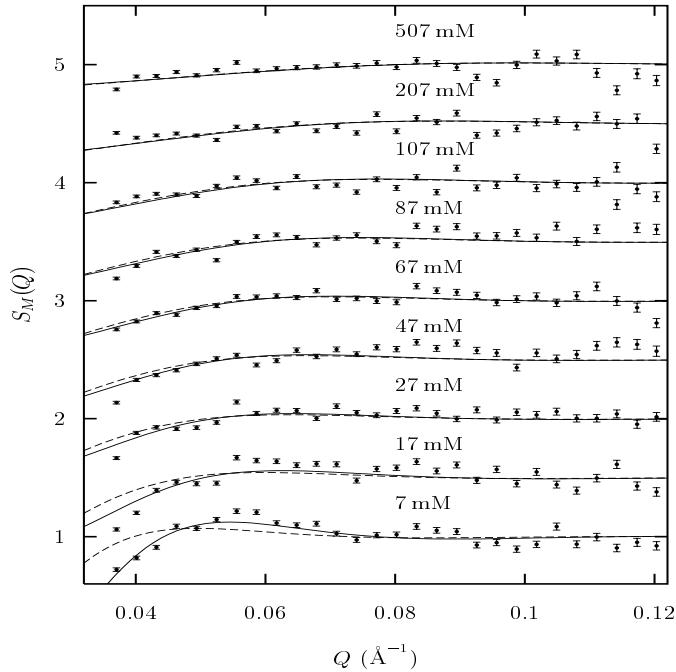} }
\caption{Comparison between the measured 
  structure factors $S_M(Q)$ for the $\beta$LG at pH=2.3 and
  concentration 10~$\rm g \, L^{- 1}$ in different ionic strength
  conditions (as indicated above each curve).  The best fit lines
  resulting from the simultaneous analysis of the corresponding SAXS
  curves (Fig.~\ref{figure1}) using the $0^{\rm th}$-order (dashed)
  and $1^{\rm st}$-order (solid) approximations of the pair
  correlation functions are reported.  Data for $Q>0.12 \An^{-1}$ are
  not shown for clarity.  }
\label{figure2}
\end{figure}

In Fig.~\ref{figure2} the theoretical results for the average
structure factor $S_{M}(Q)$ are shown along with the experimental
data.  While at high $I_S$ (i.e. at weak effective interactions) the
two approximations are practically undistinguishable, for $I_S \leq 27
\, \mbox{mM}$ the $1^{\rm st}$-order results outplay the $0^{\rm
  th}$-order ones, mainly in the low- $Q$ region.

\begin{figure}[tbp]
\centerline{\epsfxsize=3.5truein \epsfysize=3.5truein
\epsffile{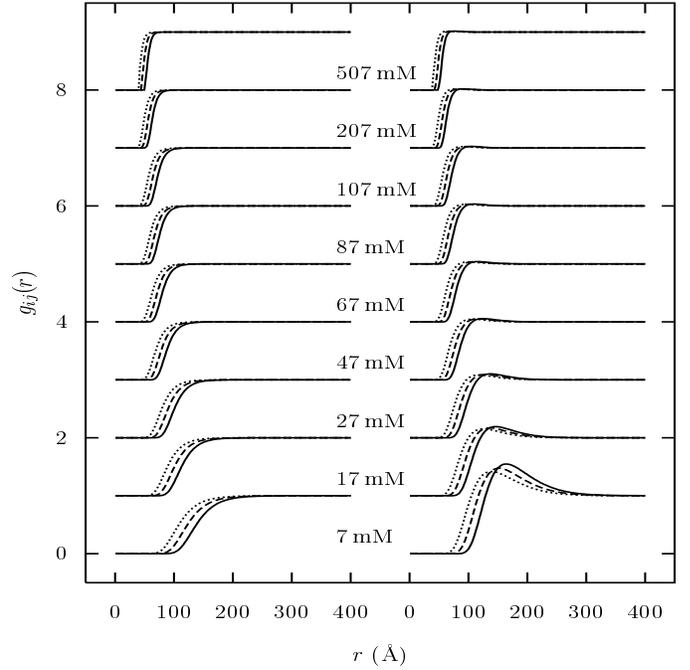} }
\caption{Partial correlation functions
  $g_{ij}(r)$ resulting from the simultaneous analysis of the nine
  SAXS curves of Fig.~\ref{figure1} (the ionic strength, $I_S$, is
  indicated near each set of curves) by applying the $0^{\rm
    th}$-order (left column) and $1^{\rm st}$-order (right column)
  approximation in the density expansion of the mean-force potential.
  Depicted are the monomer-monomer, $g_{11}(r)$ (dotted lines), the
  monomer- dimer $g_{12}(r)$ (dashed lines) and the dimer-dimer
  $g_{22}(r)$ (solid line) correlation functions.  }
\label{figure3}
\end{figure}

A more transparent comparison between the two approximations is
carried out in Fig.~\ref{figure3} at the level of RDFs. As $I_S$
decreases, the $1^{\rm st}$-order $g_{ij}(r)$ ($i,j=1,2$) become
strongly different from the $0^{\rm th}$-order ones, exhibiting a peak
of increasing height. In terms of potentials of mean force, 
 $g_{ij}(r)>1$ in some regions (mainly for $I_S \leq 27 \,
\mbox{mM}$) implies that $W_{ij}(r)<0$, although $u_{ij}(r)$ always
remains positive. The first-order correction $\omega_{ij}^{(1)}(r)n$
therefore corresponds to an {\it attractive} contribution, due to an
``osmotic depletion'' effect
%\cite{Asakura54}
(\Asakura54) exerted on two given macroparticles by the remaining
ones.  This many-body effect is clearly lacking in the $0^{\rm
  th}$-order approximation, as depicted in Fig.~\ref{figure3}.  Depletion
forces arise when two protein molecules are close together. In this
case the pressure exerted on these molecules by all other
macroparticles becomes anisotropic, leading to a strong indirect
protein-protein attraction, even though all direct interactions are
repulsive.

It is worth stressing that the behavior of the $1^{\rm st}$-order
$g_{ij}(r)$ at low ionic strength could be reproduced even by the
$0^{\rm th}$-order approximation, but only at the cost of adding some
unnecessary, 
and somewhat misleading, density-dependent attractive term to the
direct pair potentials. Our model, based only on
the physically sound repulsive part of the DLVO potential, 
turns out to be rather accurate
for the purposes of the present paper.
We have also
performed some calculations including a Hamaker term into our
perturbative scheme, without finding any significative change in the
$1^{\rm st}$-order results with respect to the previous ones.

The $1^{\rm st}$-order RDFs shown in Fig.~\ref{figure3} are
undoubtedly correctly shaped, although the peak heights might be
modified by the neglected second- and higher-order corrections to the
potentials of mean force. Unfortunately, an estimate for the magnitude
of the successive perturbative terms (depending on both concentration
and charge of the protein molecules) is 
a far more complicate task and goes beyond the scope of the present paper.
Since the resulting protein charges (see Table I) are relatively large,
 it is reasonable to expect that the contribution of the higher-order
 terms might be appreciable. As the protein concentration
increases, this correction becomes more and more significant, and eventually 
the rather good performance of our $1^{\rm st}$-order approximation must break down.

Since a direct computation of even the second order corrections 
demands a high computational effort, the accuracy of the
$1^{\rm st}$-order approximation may alternatively be investigated by checking
our RDF results against exact Monte Carlo or
molecular dynamics simulation data relevant to the same model.
A simpler indication about the limits of validity of our scheme
 may come from a systematic comparison with integral-equation 
predictions based upon more accurate closures. One could use, for
 instance, the multi-component version of the ``rescaled MSA'' approach
(Ruiz-Estrada et al., 1990), which has the advantage of being
nearly fully analytical. On the other hand, if more accurate results 
are required, then the Rogers-Young closure (\Rog84)is preferable for
our potential, but in this case the corresponding integral equations
 must be solved numerically. We have planned some investigations
in this sense, and their results will be reported elsewhere. 
However, we believe that, at the considered protein concentration,
the $1^{\rm st}$-order approximation does yield the correct trend of
the RDFs. It is our opinion that the inclusion of the neglected terms
 cannot alter the qualitative (or semiquantitative) picture of
 $\beta$LG interactions
 supported by our model, even if slightly different values for
 the best-fit parameters should be expected.

The parameter values resulting from the global best-fit procedure,
using the $0^{\rm th}$-order and $1^{\rm st}$-order approximations,
are reported in Tabs.~\ref{fitpar} and \ref{nofitpar}.

 The improved quality of the fit corresponding to the first-order
 approximation can clearly be appreciated by comparing not only the global
 $\chi^2$ value (Table I), but above all the partial 
${\bar \chi^2_\pedexp}$ ones (Table II), in particular for
 $I_S \leq 27$ mM. Although the change of global $\chi^2$  is not so large, if
one considers the relative variation of the ${\bar \chi^2_\pedexp}$'s (last
column of Table II), then the improvement is rather evident for the
 low ionic strength samples, while it becomes less and less important
 with increasing ionic strength. The proposed method is able to
 improve the goodness of the fit by about $43 \%$ for the first sample 
(where the interference peak is more pronounced). The decrease of the
 relative variation, as
 the ionic strength increases, is in agreement with the expected 
progressive weakening of protein-protein repulsions.

  Note that the values of both fitting parameters, i.e. $Z_1$ and $\Delta
G_{\mathrm{nel}}$, turn out to be very similar for both
approximations.  The scaling factors, $\kappa_\pedexp$, and the flat
backgrounds, $B_\pedexp$, are also similar for all samples and for
both approximations, confirming that no other effects, like
denaturation or larger aggregation, are really present.

\section{Conclusions}

In this paper we have presented a novel methodological approach to the
study of protein-protein interactions using SAXS techniques.  Our work
builds up upon a previous investigation by some of us
%\cite{Baldini99}
(\Baldini99).

As widely discussed by \Baldini99, the structural properties of
$\beta$LG in acidic solution, studied by light and X-ray scattering
over a wide range of ionic strength and concentration, are consistent
with the existence of monomers and dimers, and cannot be ascribed to a
 denaturation process.

Since the form factors of both the species are easily known, the
so-called ``measured'' or average structure factor $S_M(Q)$ can be
obtained from the ratio between experimental intensity and
average form factor $P(Q)$ at a certain monomer fraction 
$x_1$. 
$S_M(Q)$ is related to the protein-protein effective interactions.  
Short-range  attractive interactions like
 hydrogen bonds, responsible
of the dimer formation and strongly depending on the monomer-monomer
orientation, are taken into account using a quasi-chemical description
of the thermodynamic equilibrium between monomer and dimer forms of
$\beta$LG.  Thus, in addition to the hard core repulsions, the
effective potentials of mean force only describe long-range 
monomer-monomer, monomer-dimer and dimer-dimer electrostatic repulsions,
 which can be reduced to their orientational averages, depending only 
on the intermolecular distance $r$.

In the work by \Baldini99 all long-range protein-protein forces were
neglected, because the measured SAXS intensity was spanning a $Q$-range
 where such interactions are essentially negligible. On the
contrary, we have explicitly addressed this issue in the present work.
To this aim, i) we have extended the range of measured intensities to
lower $Q$ values in order to experimentally probe these long-range
interactions, and ii) we have proposed a simple but efficient
perturbative scheme, whose first terms are able to yield reasonably
accurate RDFs for dilute or moderately concentrate solutions of
globular proteins, with a rather little computational effort. In particular,
we have explicitly computed the $0^{\rm th}- $ and $1^{\rm st}$-order
approximations and compared their results.

The improvement in the quality of the fit for $S_M(Q)$,
 obtained with the first-order correction for the potentials
of mean force corresponding to the RDFs, with respect to the standard
zero-density approximation, is particularly visible at low ionic strength,
where Coulomb repulsions are poorly screened. In this case, the
new representation of the RDFs is able to reproduce the interference
 peak present in the experimental $S_M(Q)$, whereas the 
commonly used zero-density approximation turns out to be 
quite inadequate at low ionic strength.

Finally, two points are particularly noteworthy.

First, the adopted model allows a simultaneous fit of nine SAS curves
with only two free parameters, independent of the ionic strength,
i.e., the non-electrostatic dissociation free energy and the monomer
charge. This finding means that our simple interaction model is
already able to describe the main structural features of the examined
$\beta$LG solutions. Satisfactory results obtained by many other
structural studies on colloidal or protein solutions, based upon
similar very simplified models
%\cite{Wagner91,Krause91,DAguann92a,DAguanno92b,Nagel93,Wander94}
(\Wagner91; \Krause91; \DAguann92a; \DAguanno92b; \Nagel93; \Wander94),
suggest 
that the use of  very refined potentials, containing a large
number of different contributions,  is often unnecessary, at least
at the first stages of a research. 
Using sophisticated
interaction models may even be a nonsense, when coupled with a
simultaneous very rough treatment of the correlation functions, as is
often the case with the widely employed $0^{\rm th}$-order
approximation, in spite of the fact that the introduction 
of a larger number of parameters can clearly
improve the actual fitting of the data.
Moreover, we have pointed out that, even in models with purely 
repulsive interactions, attractive effects (due to ``osmotic depletion'')
are predicted by every sufficiently accurate theory. On the contrary,
within the zero-density approximation for the RDFs, the
same attractive effects may be reproduced only at the cost of adding
artificial contributions to the potentials.

Second, the proposed $1^{\rm st}$-order approximation to the RDFs is
really able to yield accurate predictions for the average structure
factor of weakly-concentrated protein solutions, in a rather simple
but physically sound way.  It is worth stressing that the underlying
calculation scheme is not restricted to the particular model
considered in this paper, but may be easily applied to different
spherically symmetric potentials. 
Although the limit of validity of
the $1^{\rm st}$- order approximation is still an open question,
which we are planning to investigate in future work,
we think that it may represent a new useful tool for the analysis of
experimental SAS data of globular protein solutions, when their
concentration is not too high and the strength of their interaction
forces is not too large.  When these two conditions fail, then it is
unavoidable to compute the correlation functions by exploiting some
more powerful method from the statistical mechanical theory of liquids
%\cite{Hansen86}.
(\Hansen86).  We hope, however, that this paper will stimulate the
application of the proposed $1^{\rm st}$-order approximation to
different sets of experimental data on proteins, as well as new
theoretical work on the quality and limit of this calculation scheme.

\section*{Acknowledgement}

This work has been partially supported by the grant for the Advanced
Research Project on Protein Crystallization ``Procry'' from the italian
Istituto Nazionale di Fisica della Materia (INFM). We also thank
 Bruno D'Aguanno and Giorgio Pastore for useful discussions.

\appendix

\section{Calculation of protein form 
  factors}

In detail, the scattering particle is assumed to be homogeneous and
its size and shape are described by the function $s({\bf r})$, which
gives the probability that the point ${\bf r}\equiv (r,\omega _r)$
(where $\omega _r$ indicates the polar angles $\alpha _r$ and $\beta
_r$) lies within the particle. For compact particles, like globular
proteins, this function can be written in terms of a unique two-
dimensional angular shape function $%
{\cal F}(\omega _r)$, as
\begin{equation}
s({\bf r})=\left\{
\begin{array}{ll}
1 & \quad \mbox{$r \leq {\cal F}(\omega_r)$} 
\\
\exp \{-[r-{\cal F}(\omega _r)]^2/2\sigma 
^2\} & \quad
\mbox{$r
> {\cal F}(\omega_r)$}
\end{array}
\right. \label{srdef}
\end{equation}
where $\sigma $ is the width of the gaussian that accounts for the
particle surface mobility (\Svergun98).  The shape function ${\cal
  F}(\omega _r)$ is evaluated by fixing the axis origin on the mean
value of the atomic coordinates and running over each atom $m$ and
taking the maximum distance $r$ between the origin and the
intersection, if any, of the van der Waals sphere centered in $m$ with
the direction $\omega _r$. Assuming homogeneous particles belonging to
species $i $, $M_i$ random points are generated from polar
coordinates. The sampling is made for the variables $\alpha _r$, $\cos
\beta _r$ and $r^3$ in the ranges $%
[0,2\pi ]$, $[-1,1]$ and $[0,r_{max}^3]$, respectively. Following
Eq.~\ref {srdef}, if $r \leq {\cal F}(\omega _{r})$, the point is
accepted, otherwise the probability ${\cal P}=\exp \{- [r-{\cal
  F}(\omega _{r})]^2/2\sigma ^2\}$ is calculated. A
random number $y$ between $0$ and $%
1$ is extracted and if $y<{\cal P}$ the point is accepted, otherwise
is rejected. The $p_i^{(1)}(r)$ histogram is then determined by taking
into account the distances between the $M_i$
points and the centre, while the $%
p_i^{(2)}(r)$ histogram depends on the distances between all possible
pairs of $M_i$ points,
\begin{eqnarray}
p_i^{(1)}(r) &=&\frac 1{\Delta 
rM_i}\sum_{n=1}^{M_i}H(\Delta r/2-|r-r_n|),
\label{p1risto} \\
p_i^{(2)}(r) &=&\frac 2{\Delta rM_i(M_i-
1)}\sum_{n=1}^{M_i-1}%
\sum_{m=n+1}^{M_i}H(\Delta r/2-|r-r_{nm}|), 
\label{p2risto}
\end{eqnarray}
where $\Delta r$ is the grid amplitude in the space of radial
distance, $r_n$ the distance between the centre and the $n$- th point.
Here $r_{nm}$ is the distance between the points $n$ and $m$, and
$H(x)$ is the
Heaviside step function ($%
H(x)=0$ if $x<0$ and $H(x)=1$ if $x\geq 0$).  The number of random
scattering centres was $M_i=2000$, the grid size was $\Delta r = 1
\An$, while the width of the surface mobility was fixed to $\sigma=2 {\ {\rm
    \AA}}$.

\section{First-order perturbative 
  corrections}

In the density expansion of the potentials of mean force $W_{ij}\left(
  r\right) $
\begin{equation}
-\beta W_{ij}\left( r\right) = -\beta u_{ij}\left( r\right) 
+\omega _{ij}^{(1)}(r) n +
\omega _{ij}^{(2)}(r) n^2 + \ldots ,
\end{equation}

\noindent the {\it exact} power coefficients $\omega _{ij}^{(k)}(r)$ (
$k=1,2,\ldots$) can be computed by using standard diagrammatic techniques
%\cite{Meeron58}
(\Meeron58), which yield the results in terms of appropriate multi-
dimensional integrals of products of Mayer functions
\begin{equation}
f_{ij}\left(\ r\right) =\exp \left[ -\beta 
u_{ij}\left( r\right) \right] -1
\end{equation}

Within our approximation, we are only required to compute the first
term, which involves a convolution and turns out to be
\begin{equation}
\omega _{ij}^{(1)}(r)=\sum_k \molratio 
\gamma
_{ij,k}^{(1)}(r)=\sum_k \molratio \int {\rm 
d}{\bf r}^{\prime }~f_{ik}\left(
r^{\prime }\right) ~f_{kj}\left( |{\bf r-
r}^{\prime }|\right) ,
\end{equation}
where $ \molratio =n_k/n$ is the molar fraction of species $k$.
%and
%%
%\begin{equation}
%\gamma_{ij,k}^{(1)}(r)=\int {\rm d}{\bf 
%r}^{\prime }~f_{ik}\left( r^{\prime
%}\right) ~f_{kj}\left( |{\bf r-r}^{\prime 
%}|\right).
%\end{equation}
%
The evaluation of the convolution integral $\gamma _{ij,k}^{(1)}(r)$
is not a difficult task in bipolar coordinates.  Integration over
angles is easily performed and $\gamma _{ij,k}^{(1)}(r)$ reduces to a
double integral, which can be written as
\begin{equation}
\gamma _{ij,k}^{(1)}(r)=\frac{2\pi 
}r\int_0^\infty dx\ \left[ xf_{ik}\left(
x\right) \right] \int_{\left| x-r\right| 
}^{x+r}dy\ [yf_{kj}(y)].
\label{convolution}
\end{equation}
We have evaluated all these $\gamma _{ij,k}^{(1)}(r)$ terms at the
points $r_i = i \Delta r$ ($i = 1,\ldots,500$), with $\Delta r = 1
\An$.  At each $r_i$ value, the double integral has been carried out
numerically, simply by using the trapezoidal rule for both $x-$ and $y$-
integration. For the $x$-integration, we have chosen as upper limit
the value $x_{\rm max} = \max ( x_{\rm cut}, R_2 + r )$, with $x_{\rm
  cut} = R_2 + 12 / \kappa_D$ (depending on the ionic strength), and
as grid size $\Delta x = x_{\rm cut} / 200$. For the $y$-integration,
$\Delta y = \Delta x$.

%%%%%%%%%%%%%%%%%%%%%%%%%%%%%% FIGURES %%%%%%%%%%%%%%%%%%%%%%%%%%%%%%%%%%
%FIG1
\vskip 2.0 cm
%FIG2
\vskip 2.0 cm
%FIG3

\newpage
%%%%%%%%%%%%%%%%%%%%%%%%%%%%%% TABLES %%%%%%%%%%%%%%%%%%%%%%%%%%%%%%

%TAB1
\begin{table}[h]
\begin{tabular}{|cccc|}
approx. & $Z_1$ & $\Delta G_{\mathrm{nel}} / 
k_B T$ & $\chi^2$\\
\hline
 $0^{\rm th}$ & $19.6 \pm 0.1$ & $14.8 \pm 
0.1$ & $10.9$ \\
 $1^{\rm st}$ & $20.0 \pm 0.2$ & $16.6 \pm 
0.1$ & $8.9$ \\
\end{tabular}
\vskip 1.0 cm
\caption{Comparison of the fitting 
parameters 
(the monomer effective charge, $Z_1$, and 
the non-electrostatic free
energy, $\Delta G_{\mathrm{nel}}$) and of 
the merit functional
$\chi^2$
resulting from the simultaneous analysis
of the nine SAXS curves of Fig.~\ref{figure1}
by applying 
the $0^{\rm th}$-order and $1^{\rm st}$-
order approximations of 
the pair correlation functions.}
\label{fitpar}
\end{table}
\vskip 1.0 cm
%TAB2  
\begin{table}[h]
\begin{tabular}{|cccccccc|}
$I_S$ & $\kappa_\pedexp$ && $B_\pedexp$ && ${\bar 
\chi^2}_\pedexp$ &&\\
(mM) & ($10^{-3}$ a.u. cm) && ($ 10^{-5} $ 
a.u.) &&&&\\
\hline
& $0^{\rm th}$ & $1^{\rm st}$ & $0^{\rm th}$ 
& $1^{\rm st}$ & $0^{\rm 
th}$ & $1^{\rm st}$ & Var $(\%)$ \\
\hline
 7 & $1.450 \pm 0.002$ & $1.478 \pm 0.002$ 
&$4.62 \pm 0.06$ &$4.48 
\pm 0.06$ &$14.2$& $ 8.0$ & $-43.7$  \\
17 & $1.424 \pm 0.002$ & $1.424 \pm 0.002$ 
&$4.73 \pm 0.05$ &$4.73 
\pm 0.05$ &$14.8$& $10.7$ &  $-27.7$  \\
27 & $1.619 \pm 0.003$ & $1.521 \pm 0.003$ 
&$4.79 \pm 0.05$ &$5.23 
\pm 0.05$ &$10.9$& $ 8.7$ &  $-20.2$  \\
47 & $1.397 \pm 0.003$ & $1.293 \pm 0.003$ 
&$3.46 \pm 0.05$ &$3.98 
\pm 0.04$ &$10.6$& $ 9.9$ &  $-6.6$  \\
67 & $1.443 \pm 0.002$ & $1.367 \pm 0.002$ 
&$3.78 \pm 0.05$ &$4.25 
\pm 0.06$ &$ 7.7$& $ 5.5$ &  $-28.6$   \\
87 & $1.405 \pm 0.003$ & $1.351 \pm 0.003$ 
&$4.18 \pm 0.06$ &$4.47 
\pm 0.07$ &$12.0$& $11.8$ &  $-1.7$  \\
107& $1.493 \pm 0.003$ & $1.450 \pm 0.002$ 
&$2.06 \pm 0.06$ &$2.30 
\pm 0.06$ &$ 9.3$& $ 8.2$ &  $-11.8$  \\
207& $1.478 \pm 0.002$ & $1.457 \pm 0.002$ 
&$4.12 \pm 0.06$ &$4.23 
\pm 0.06$ &$10.1$& $ 9.5$ &  $-5.9$   \\
507& $1.529 \pm 0.003$ & $1.518 \pm 0.003$ 
&$3.68 \pm 0.08$ &$3.73 
\pm 0.08$ &$ 8.3$& $ 8.0$ &  $-3.6$  \\
\end{tabular}
\vskip 1.0 cm
\caption{Comparison of the scaling factors, 
$\kappa_\pedexp$, the flat backgrounds, $B_\pedexp$,
and the merit functionals, ${\bar \chi^2}_\pedexp$ 
(Eq.~\ref{partial}), 
resulting from the simultaneous analysis
of the nine SAXS curves of
Fig.~\ref{figure1})
by applying 
the $0^{\rm th}$-order and $1^{\rm st}$-
order approximations of 
the pair correlation functions. 
The last entry Var ($\%$)
provides the relative variation between 
the $0^{\rm th}$-order and $1^{\rm st}$-
order approximations.
}
\label{nofitpar}
\end{table}
\newpage
%%%%%%%%%%%%%%%%%%% REFERENCES %%%%%%%%%%%%%%%%%%%%%%%%%%%%%%

\end{document}